\documentclass[aps,prl,superscriptaddress,twocolumn,showpacs,floatfix]{revtex4-1}
\usepackage{graphicx}
\usepackage{graphicx}
\usepackage{amsmath, amsfonts, amssymb,mathrsfs}
\usepackage{bm}

\providecommand\figwidth{3.375in}
\graphicspath{{Figs/}}

\providecommand{\abs}[1]{\left\lvert#1\right\rvert}   
\providecommand{\av}[1]{\left<#1\right>}
\providecommand{\ket}[1]{\left|#1\right\rangle}

\providecommand{\bra}[1]{\left\langle#1\right|}

\renewcommand{\vec}[1]{\bm{#1}}

\providecommand{\rel}{\phi}
\providecommand{\irr}{\eta}
\providecommand{\fun}{h}
\providecommand{\Lc}{L_c}
\providecommand{\Len}{L_0}
\providecommand{\dof}{d}
\renewcommand{\section}[1]{\emph{#1}}

\begin{document}
\title{Scaling and localization lengths of a topologically disordered system}
\author{Jacob J.\ Krich}
\affiliation{Harvard University Center for the Environment, Cambridge, MA 02138}
\affiliation{Department of Chemistry and Chemical Biology, Harvard University, Cambridge, MA 02138}
\author{Al\'an Aspuru-Guzik}
\affiliation{Department of Chemistry and Chemical Biology, Harvard University, Cambridge, MA 02138}
\date{January 19, 2011}

\begin{abstract}
    We consider a noninteracting disordered system designed to model particle diffusion, relaxation in glasses, and impurity bands of semiconductors. Disorder originates in the random spatial distribution of sites. We find strong numerical evidence that this model displays the same universal behavior as the standard Anderson model. We use finite-size-scaling to find the localization length as a function of energy and density, including localized states away from the delocalization transition. Results at many energies all fit onto the same universal scaling curve.
\end{abstract}
\pacs{xxx}
\maketitle

The disorder-induced transition from extended to localized states in non-interacting quantum systems has been a rich source physics insight for over fifty years \cite{Anderson58}. It is relevant for a broad range of transport properties \cite{Lee85}, glass formation \cite{Amir10}, conductivity of composites \cite{Ambrosetti10}, random walks \cite{*[{}] [{, p. 29.}] Mulken11}, as well as for non-radiative recombination in intermediate-band photovoltaics \cite{Luque06,*Lopez11}. Scanning-tunneling and Bose-Einstein condensate experiments are increasingly able to probe localization properties directly \cite{Morgenstern02,*Hashimoto08,*Richardella10,*Clement06,*Billy08}. The delocalization transition is usually studied using the standard Anderson model, which considers a non-interacting tight-binding lattice with uniform nearest-neighbor coupling and random on-site energies \cite{Anderson58}. For systems in which the disorder originates in the random configuration of the sites rather than, e.g., random local fields, it is better to consider the so-called topologically disordered or Lifshitz model, in which sites are distributed randomly in space with no on-site energies and all pairs of sites are connected by hopping terms with amplitude exponentially decaying with the distance between them \cite{Lifshitz65}. As the density of sites increases, a localization/delocalization transition occurs, just as in the standard Anderson model. The density of states \cite{Lifshitz65,Matsubara61,Ching82,*Logan85,*Logan86,Shklovskii84,Mezard99,Amir08,Amir10} and localization properties \cite{Ching82,*Logan85,*Logan86,Priour10,Amir10} of this model have received much attention. Here we obtain the localization lengths of this model quantitatively as a function of wavefunction energy and site density. We adapt the finite-size scaling method, which has been successfully applied to the standard Anderson model \cite{MacKinnon81,MacKinnon83,Slevin99,*Slevin03,Rodriguez10,Markos06}, and give strong numerical evidence that the topologically disordered model displays the same universal behavior.

\section{Model}
We consider particles confined to identical lattice sites distributed randomly in space, with density $\rho'$. Particles can hop between sites with an exponentially decaying hopping coefficient. The Hamiltonian is
\begin{align}\label{eq:H}
  H=-\frac{V_0}{2} \sum_{n\neq m} e^{-r_{nm}/a_B^*} \ket{n}\bra{m}
\end{align}
where $\ket{n}$ are the site wavefunctions, $r_{nm}$ is the distance between sites $n$ and $m$, $a_B^*$ is an effective Bohr radius giving the decay of the wavefunctions, $V_0$ is an overall energy scale, and the sum runs over all pairs of sites. This model is called topologically disordered because there is no ordered structure describing which lattice sites are strongly coupled to each other. This model can describe the Hamiltonian of impurities with hydrogenic wavefunctions in a semiconductor, where the hopping originates in overlaps of their effective atomic wavefunctions. The dimensionless density is $\rho\equiv\rho' a_B^{*3}$. Experimentally, such systems are found to have a metal-insulator transition at $\rho^{1/3}\approx0.26$ \cite{Edwards78}. If the diagonal components of $H$ are set such that each column sums to zero, the model can describe glass dynamics \cite{Amir08,Amir10} and continuous time quantum walks \cite{Mulken11}, but many properties are similar.

\begin{figure}
\includegraphics[width=\figwidth]{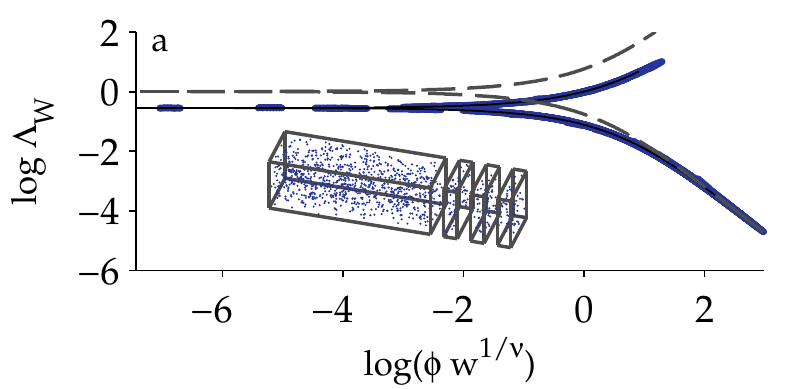}
\includegraphics[width=\figwidth]{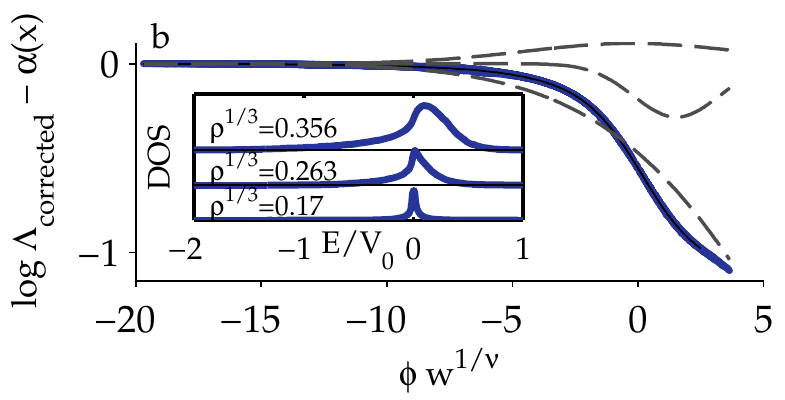}
  \caption{\label{fig:DOS}
  Scaling function for data with $w\geq25$, including one correction to scaling $\irr$ with $y=2.7$. The thin solid line is the fit. Statistical errors in $\Lambda_w$ are smaller than point sizes. There are 5829 data points and 893 fit parameters. The normalized $\chi^2$ statistic per degree of freedom $\dof$ is 7.5. 70\% of the data points are within error of the fit. In (a), the dashed line shows the asymptote $\alpha(x)=\nu\sinh^{-1}(x/2)$, showing good agreement in the strongly localized regime. In (b), the relevant function $h_0(x)$ minus the asymptotic $\alpha(x)$ is plotted. Dashed lines show the three Gaussians from the fit of $\delta$.
  (inset to a) A wire of width $w$ is constructed by sequentially adding slices (shown at right) to an existing wire. Each slice has length $\Len\geq\Lc$ and has randomly distributed particles of the chosen density.
  (inset to b) Density of states for three site densities, calculated with $1500$ sites and 100 realizations of disorder. Curves offset for clarity. At high density, the DOS becomes asymmetric.
 }
\end{figure}

The density of states (DOS) is shown in the inset of Fig.\ \ref{fig:DOS}b. The DOS is well-understood in the low-density \cite{Amir10} and high-density, low energy \cite{Matsubara61,Mezard99} limits. 

\section{Localization Lengths from Quasi-1D Scaling}
The localization length $\lambda$ of a localized eigenstate is determined by the asymptotic decay of the wavefunction $\psi\sim e^{- \abs{\vec{r}-\vec{r}_0}/\lambda}$, where $\vec{r}_0$ is some location of high wavefunction amplitude. Our goal is to find the localization length as a function of $\rho$ and $E$. Finite-size scaling techniques allow us to study computationally tractable small systems and systematically extrapolate to results for the true infinite system.
We assume, and find strong numerical evidence to confirm, that the topologically disordered system is controlled by the same fixed point as the standard Anderson model. The critical exponent and other critical parameters of this fixed point have been well-determined by previous studies \cite{MacKinnon94,Slevin99,Slevin03,Rodriguez10}, and we will use these results to aid our study of the localization lengths. While studies interested in the critical exponents have focused on systems close to the localization transition, we are interested in localization lengths across the range of energies and densities relevant to experiments, which includes not only values close to the transition but also ones in the strongly localized regime.  As we move away from the critical point, corrections to scaling should become more important, and we test our results' sensitivity to these corrections.

The quasi-1D scaling method was introduced for the standard Anderson model by MacKinnon and Kramer, who considered several wires of varying widths $w$ and long lengths \cite{MacKinnon81}.  
We adapt the recursive Green function technique to find the localization length $\lambda_w$ for equivalent wires in our system, see inset to Fig.\ \ref{fig:DOS}a \cite{MacKinnon81,MacKinnon83}.
The key idea is to divide a long wire into slices, where the sites in each slice are directly coupled only to each other and to sites in the immediately adjacent slices. This is not strictly possible for the system of Eq.\ \ref{eq:H}, as all sites are directly coupled. We take, however, a cutoff $\Lc$ and set $\exp(-r/a_B^*)\rightarrow0$ for $r\geq\Lc$. 
We start with a single slice of width $w$, length $\Len\geq\Lc$ and periodic boundary conditions in two dimensions. We recursively add slices adjacent to the wire and find the portion of the Green function that connects any site in slice 1 to any site in slice $N$ at fixed energy $E$. We then use the standard method to determine the localization length $\lambda_w$ of the long wire at energy $E$ \cite{MacKinnon81,MacKinnon83,Markos06}. The width $w$ of each wire must be large enough that there is a sufficient probability that each slice will have sites within $\Lc$ of sites in the neighboring slices, otherwise the wire is disconnected \cite{Taylor89}.
Due to the varying number of particles in each slice and the varying off-diagonal matrix elements, the transfer matrix method (reviewed in Ref.\ \onlinecite{Markos06}) cannot be used.

The statistical error 
in $\lambda_w$ can be estimated by assuming that the estimate of $\lambda_w$ after $N$ slices is the mean of $N$ independent and identically distributed samples chosen from a normal distribution, so the sampling error goes down as $N^{-1/2}$ \cite{Markos06}. This is rigorously proved for the standard Anderson model with Oseledec's theorem \cite{Markos06}, and we assume that similar statistics hold here. We choose a maximum error of 1\%, which requires $10^4$ to $10^6$ slices, depending on the parameters.

After determining $\lambda_w$ for a range of widths $w$, MacKinnon and Kramer (and many subsequent authors) use the one-parameter scaling theory \cite{Abrahams79} to extrapolate to infinite size systems.
The scaling can be expressed in terms of the relevant variable $\rel$, with $\rel$ greater (less) than zero for extended (localized) states. Then the dimensionless quantity $\Lambda_w\equiv\lambda_w/w$ is a universal function of $\rel$ and $w$, with critical exponent $\nu$, \cite{Slevin99}
\begin{align}\label{eq:fScale}
  \Lambda_w=f[\rel(\rho,E) w^{1/\nu}],
\end{align}
The correlation length is $\xi=\abs{\rel}^{-\nu}$. 

The hypothesized universality of the metal-insulator transition with time reversal symmetry has been confirmed in a variety of different models \cite{Slevin99,Markos06}.  If we assume that this topologically disordered model is controlled by the same fixed point as the standard Anderson model, then we expect Eq.\ \ref{eq:fScale} to hold in our system, with the same $f(x)$. Previous work has expanded $f(x)$ in polynomials in order to fit the data \cite{Slevin99,Slevin03,Rodriguez10}. We have not found this to be a successful strategy, at least in part because the underlying functions are strongly non-polynomial away from the transition region (see below). In general in such fits, there is a large number of parameters \cite{Slevin99,Rodriguez10}, so it is helpful to constrain as many of them as possible.
We assume universality and use four pieces of information from previous work: $\nu=1.58$ and $f(0)=0.576$, as shown for the standard Anderson model \cite{Slevin99}. Further, we have asymptotic limits  $f(x\rightarrow\pm\infty)=\abs{x}^{\pm\nu}$  \cite{MacKinnon83}. We choose to fit to $\log\Lambda_w=\log f(\rel w^{1/\nu})\equiv\fun_0(\rel w^{1/\nu})$. Then we have $\fun_0(0)=-0.55$ and $\fun_0(x\rightarrow\pm\infty)=\pm\nu\log(\pm x)$.
The asymptotic limits are obeyed by the function $\alpha(x)=\nu \sinh^{-1}(x/2)$. We then fit our data to
 $ \fun_0(x)=\alpha(x)+\delta(x)$ 
where $\delta(0)=-0.55$ and $\delta(\abs{x}\rightarrow\infty)=0$.

The scaling form, Eq.\ \ref{eq:fScale}, applies only for $w$ sufficiently large. 
At small $w$, corrections to scaling modify Eq.\ \ref{eq:fScale}, and they have proven essential for accurate determination of $\nu$ and $\Lambda_c$ \cite{Slevin99}. Corrections to scaling require introduction of some number of irrelevant variables $\irr_i(E,\rho)$, which have no effect on the scaling in the limit $w\rightarrow\infty$. We illustrate with only one irrelevant variable $\irr$, but more are easily added. We rewrite our scaling equation as
  $\log \Lambda_w=\fun(\rel w^{1/\nu},\irr w^{-\abs{y}})$,
where $y$ is another critical exponent. We expand $\fun$ in powers of $\irr w^{-\abs{y}}$,
\begin{align}\label{eq:h1irrexpanded}
 \log \Lambda_w=\sum_{m=0} \irr^m w^{-m\abs{y}}\fun_m(\rel w^{1/\nu}),
\end{align}
where $\fun_0$ is the limiting one-parameter scaling function. It is often sufficient to keep only $m=0,1$ in Eq.\ \ref{eq:h1irrexpanded}, \cite{Slevin99,Slevin03} which we will do here. We then define $\log \Lambda_\text{corrected}\equiv\log\Lambda_w-\irr w^{-\abs{y}}\fun_1(\rel w^{1/\nu})$.

We are interested in finding the localization length at many values of $E$, unlike the usual choice of $E=0$. The Green functions in the determination of $\lambda_w$ must be evaluated separately for each choice of $E$, increasing the computational requirements. We choose 14 values of density and 40 values of energy satisfying $-2\leq E/V_0\leq 0.5$, which are focused around the relevant energies for the metal insulator transition near $\rho^{1/3}\approx0.26$. We do not study $E=0$ because isolated sites always produce eigenvalues with $E=0$, causing divergences in the Green functions. We take $\Len=15 a_B^*$ and $\Lc=\min[\Len,w/2]$.  At each $\rho$, we take $w$ from $22 a_B^*$ in increments of $3 a_B^*$ up to the largest system size we find computationally tractable. For the 14 densities, the maximum $w/a_B^*$ studied are $(121, 103, 88, 79, 70, 61, 55, 52, 46, 43, 40, 37, 34, 31)$, from lowest to highest density. Previous studies of mobility edges and critical exponents have performed separate scaling at each $E$ \cite{Bulka85,*Kramer90}. In this work, all of the data collapse onto a single scaling curve, see Fig.\ \ref{fig:DOS}, showing universality independent of energy. These calculations used approximately 24 CPU-years of computing time.

We approximate $\delta(x)$ as a sum of three Gaussians, fixed to have $\delta(0)=-0.55$, with eight free parameters. We use a direct search algorithm 
as implemented in MATLAB's \texttt{fminsearch} function to find the least-squares optimal choice of these eight parameters.  Within the search loop, for each proposed $\delta(x)$, we find $\rel(\rho,E)$ (and possibly $\irr(\rho,E)$) independently at each value of $(\rho,E)$, giving several hundred additional parameters. It would be preferable to have a functional form for $\rel(\rho,E)$, but we did not find a parametrization that permitted accurate fits. To ensure that these fits are not overdetermined, we include in our dataset only $(\rho,E)$ with at least 4 values of $w$; the average number of values of $w$ at each $(\rho,E)$ is 13. 
In fits including corrections to scaling, we approximate $h_1(x)$ as a second order polynomial with $h_1(0)=1$, which fixes the scale of $\irr$.  

Fig.\ \ref{fig:DOS}a shows the scaled data, with one correction to scaling.  The upper branch shows the extended states ($\Lambda\rightarrow\infty$ as $w\rightarrow\infty$) and the lower branch shows localized states. The scaling is excellent, which justifies the use of $\nu$ and $\Lambda_c$ from the standard Anderson model studies. The consistency of $\Lambda_c$ with previous work indicates that the lattice spacing in the standard Anderson model is equivalent to $a_B^*$ in this model. The extended state curve does not approach its $\rel\rightarrow\infty$ asymptote due to a lack of data in the computationally demanding high density regime.
Fig.\ \ref{fig:DOS}b shows the same fit with $\alpha$ and the irrelevant corrections subtracted. The dashed lines show the constituent Gaussians in $\delta$, which are not well-constrained in the fits. See supplementary information for other fits.
Fig.\ \ref{fig:xi} shows the resultant correlation length $\xi(\rho,E)=\abs{\rel}^{-\nu}$.
At each $\rho$, the correlation length falls smoothly as $E$ moves away from the delocalization transition, just as we expect. These results give confidence that the $\rel(\rho,E)$ are not simply arbitrary parameters that happen to produce good scaling fits but rather are determined by the underlying physics. We find, however, that $\rel(\rho,E)$ (and thus $\xi(\rho,E)$) is quantitatively determined only for localized states.

To assess the quality of the scaling curves produced, we consider the $\chi^2$ statistic. 
Previous high accuracy numerical works on the standard Anderson model have evaluated their fits with the quality of fit $Q$ (also known as the p-value) from the $\chi^2$ statistic and the number of fitting parameters $N_P$ \cite{Slevin99,Slevin03}. All fits shown here have $Q=0$, indicating that statistical fluctuations of the numerical procedure alone are insufficient to explain the deviations of the data from the model.  This is not, however, surprising. Our study has several thousand degrees of freedom, $\dof=N_D-N_P-1$, where $N_D$ is the number of data points, which gives it statistical power to detect relatively small deviations between the model and the data. In the true scaling function, $\delta(x)$ is not actually a sum of three Gaussians, and we are sensitive to the deviations between the true $\delta(x)$ and its model. We should be able to add more Gaussians to $\delta$ to better approach the universal function, but fitting becomes difficult. If we are not entirely interested in the exact shape of $\delta(x)$, then this deviation is not a concern.  
If we limit our dataset to contain only $(\rho,E)$ close to the delocalization transition, as in previous studies of the standard Anderson model, we can obtain good values of $Q$; this occurs because the number of data points is significantly reduced and the scaling function $h(x)$ can be accurately approximated as a polynomial.

\begin{figure}
  \includegraphics[width=\figwidth]{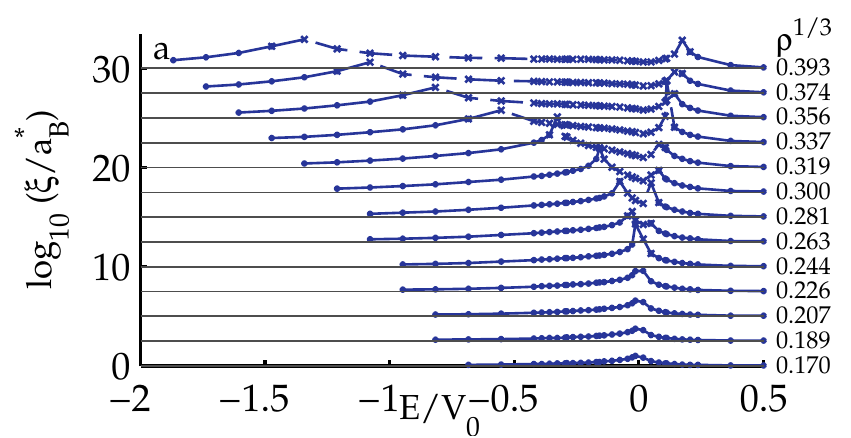}
    \includegraphics[width=\figwidth]{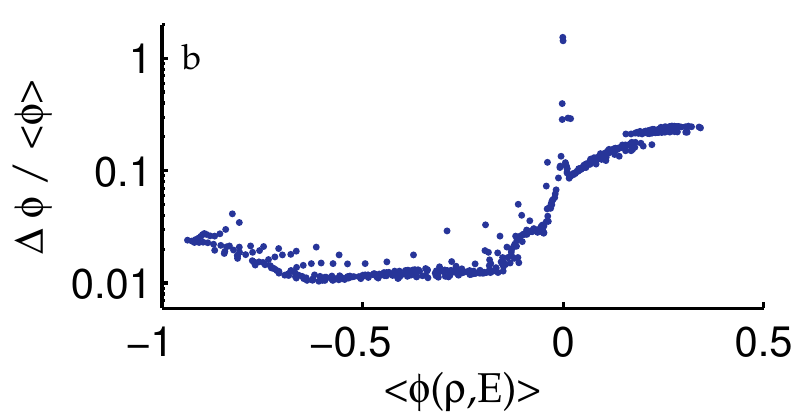}
  \caption{\label{fig:xi}
  (a) Correlation lengths $\xi(\rho,E)$ from the fit of Fig.\ \ref{fig:DOS}, for each of the 13 densities which produced enough data to be studied, offset for clarity. Solid circles mark localized states and crosses mark extended states. Solid and dashed lines are guides to the eye.
  As expected, the correlation length increases smoothly as the mobility edge is approached from either side. This figure shows the choice of energies for study, which are focused on the area of interest for the critical density, near $\rho^{1/3}=0.24$.  The mobility edge is asymmetric just as is the density of states, with the upper mobility edge closer to $E=0$ than the lower mobility edge.  The values of ($\rho$,$E$) in the lower-left corner of the figure have extremely low density of states, contributing less than 0.1\% of total states, so are excluded.
  (b) Deviations in fitted values of the scaling variable $\rel(\rho,E)$ from 92 different fits with and without corrections to scaling and with the smallest value of $w$ taken to be 22, 25, or 28 (in units of $a_B^*$). Localized states with $\phi<-0.05$ are quantitatively determined within 10\% by the scaling fits. See the Supplementary Information for more versions of this plot.}
\end{figure}

Since we are interested in extracting correlation lengths $\xi=\abs{\rel}^{-\nu}$, a better determination of the quality and confidence of the fits is to compare the resultant $\rel(\rho,E)$ for different fitting procedures.  We compare fits with and without corrections to scaling and with the smallest $w$ being 22, 25, or 28, in units of $a_B^*$. Depending on initial guesses, fits can arrive at a number of different local minima. We use a multistart procedure for the fitting, starting with 100 widely varying parameters for $\delta(x)$ and $h_1(x)$. We find that the ``best fits,'' judged solely by minimizing $\chi^2$, have highly oscillatory $\delta(x)$, discontinuous $\rel(\rho,E)$, large deviations from the asymptotic form even for strongly localized states, or large corrections to scaling; they generally have multiple of these features. If we exclude the fits with these four characteristics, for $w\geq25$ the best fits without (with) corrections to scaling have $\chi^2/\dof\approx 70$ $(8)$. Including the anomalous fits, we can find $\chi^2/\dof\approx31$ $(4)$. We find $\rel(\rho,E)$ from 92 different fits and, independently at each $(\rho,E)$, find the mean $\av{\rel}$ and standard deviation $\Delta \rel$, shown in Fig.\ \ref{fig:xi}b. The values of $\rel(\rho,E)$ are found to vary by less than 10\% in the localized regime, except for the points closest to the delocalization transition. For $\xi<100 a_B^*$, we can consider the localization lengths to be given quantitatively by the scaling method. Due to the lack of strongly-extended states, $\delta(x>0)$ is not well-determined, and the fits show a range of different shapes. It is then no surprise that $\rel(\rho,E)$ in the delocalized region varies widely. 
Accumulating more data in the extended regime should fix this problem.

This scaling technique quantitatively gives the localization lengths at nearly any localized $(\rho,E)$ we care to study. Application of this method to systems with on-site disorder, in addition, should shift the mobility edges ``inwards'' so that more states are localized. These localization lengths allow insight into the properties of a range of material systems, and in future work we will consider their effects on intermediate band photovoltaics.

\begin{acknowledgments}
  We acknowledge fruitful conversations with Bertrand Halperin, Kristin Javaras, Mark Winkler, Justin Song, Man-Hong Yung, Ari Turner, and Mauricio Santillana and use of Odyssey, supported by the FAS Research Computing Group. We acknowledge support of the Ziff Environmental Fellowship of the Harvard University Center for the Environment, NSF DMR-0934480, and DARPA.
\end{acknowledgments}

\bibliography{b-Si}

\newpage
\mbox{}
\newpage
\mbox{}

\begin{figure}[b]
\includegraphics[width=\figwidth]{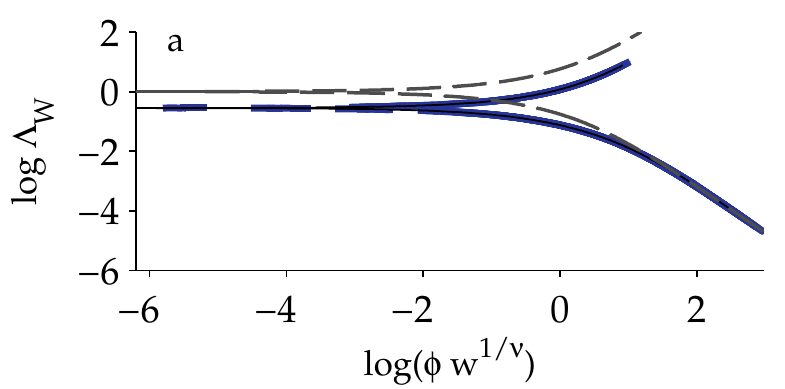}
\includegraphics[width=\figwidth]{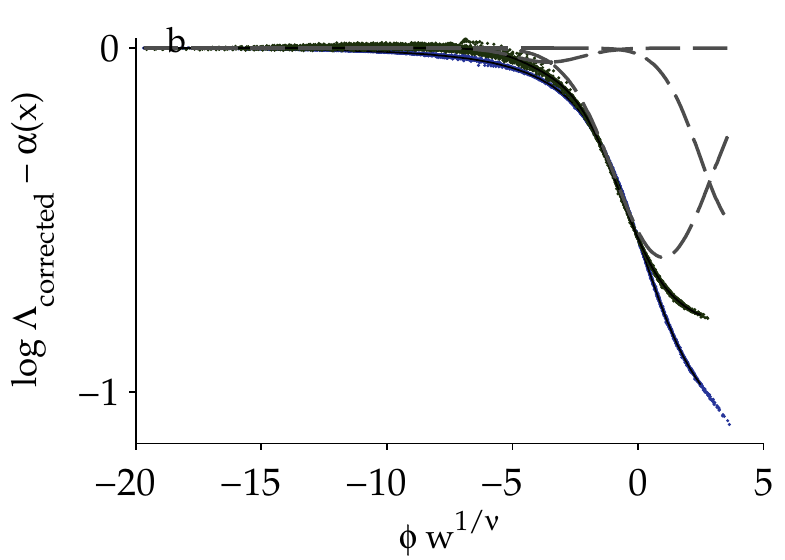}
  \caption{\label{fig:scalesupp}
  Scaling function for data with $w\geq25$, including no corrections to scaling. The scaling is very similar to that in Fig.\ 1 and produces similar $\xi$ (see Fig.\ \ref{fig:xisupp}a).
  In (b), blue points are the same as in Fig.\ 1b while green points are those of the new fit. Dashed lines show the three Gaussians from the new fit of $\delta$, which are clearly different from those in Fig.~1. There are 5829 data points and 449 fit parameters. The normalized $\chi^2$ statistic per degree of freedom $\dof$ is 71.2. 39\% of the data points are within error of the fit.
 }
 \end{figure}

\appendix
\textbf{APPENDIX: Scaling curves without corrections to scaling}\\
As discussed in the main text, several fitting procedures can be used to produce plots similar to those in Figs.\ 1 and 2. We illustrate here that the choice of fitting procedure does not significantly change the localization lengths extracted. Fig.\ \ref{fig:scalesupp} shows fits equivalent to those in Fig.\ 1, without including corrections to scaling. The scaling still looks very good, though Fig.\ \ref{fig:scalesupp}b shows that the data points do not fall on the scaling curve as precisely as in Fig.\ 1b.  The overall shape of the scaling curve deviates from that in Fig.\ 1 mostly in the extended region, $\rel>0$, which is not well determined from the data. Additionally, though the overall curve is very similar in shape to that in Fig.\ 1b, the underlying Gaussians are different, which illustrates the difficulty in converging the fits of $\delta(x)$.

 \begin{figure}
 \includegraphics[width=\figwidth]{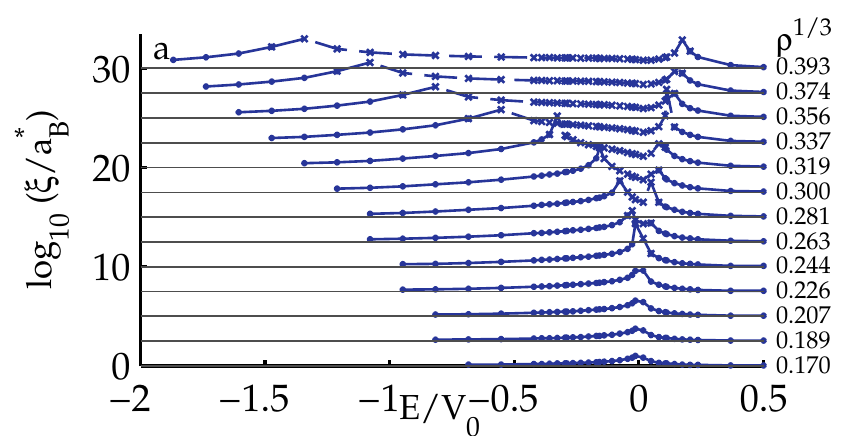}
 \includegraphics[width=\figwidth]{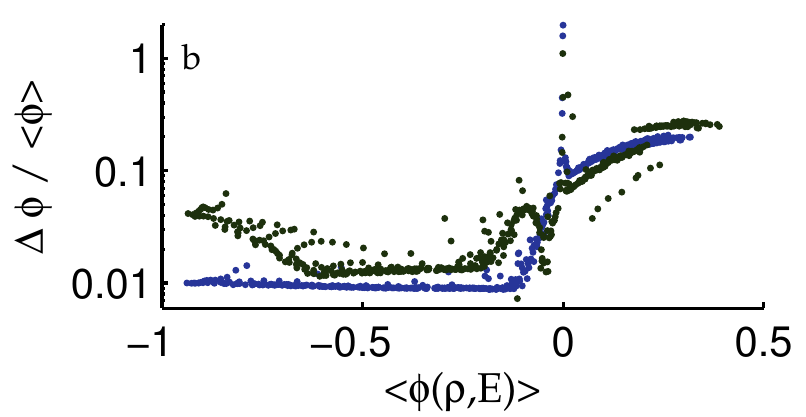}
 \caption{\label{fig:xisupp}
 (a) Correlation lengths $\xi(\rho,E)$ from the fit of Fig.\ \ref{fig:scalesupp}, for each of the 13 densities which produced enough data to be studied, offset for clarity. Plot is similar to Fig.\ 2a. Solid circles mark localized states and crosses mark extended states. Solid and dashed lines are guides to the eye.
 (b) Similar to Fig.\ 2b, deviations in the fitted values of $\rel(\rho,E)$. Blue points show deviations for 65 fits with no corrections to scaling. Green points show deviations for 27 fits with corrections to scaling.  Both include data sets with $w\geq22$, $25$, and $28$.
 }
\end{figure}

The correlation lengths $\xi$ extracted from these fits are shown in Fig.\ \ref{fig:xisupp}a. They are clearly very similar to those in Fig.\ 2a, showing that the scaling fits robustly determine $\xi(\rho,E)$, regardless of whether corrections to scaling are used.  Fig.\ 2b shows the deviations in the fitted values of $\rel$ for 92 different fits with and without corrections to scaling. 

Fits including corrections to scaling have larger disagreements in $\rel(\rho,E)$. Fig.\ \ref{fig:xisupp}b shows the deviations for 65 fits without corrections to scaling (blue) and 27 fits with corrections to scaling (green).  The fits without corrections to scaling have a smaller number of fit parameters, which seems to constrain the fits more. In any of the cases studied, the states with $\xi<100 a_b^*$ are quantitatively determined within 10\% by the fits. The consistency of the fits without corrections to scaling argues for use of that method in future studies, despite the large values of $\chi^2$ produced by ignoring the corrections to scaling.
\end{document}